\documentclass[preprint,showpacs,preprintnumbers,amsmath,amssymb,nofootinbib]{revtex4}
\usepackage[dvipdfmx]{graphicx}
\usepackage{amsmath,amssymb}
\usepackage{bm}% bold math
\usepackage{color} % For blue in-text comments and additions

\def\up{\uparrow}
\def\dwn{\downarrow}

\def\lesssim{\ \raise.3ex\hbox{$<$}\kern-0.8em\lower.7ex\hbox{$\sim$}\ }
\def\gesim{\ \raise.3ex\hbox{$>$}\kern-0.8em\lower.7ex\hbox{$\sim$}\ }
\def\bm#1{\mathbf{#1}}

\begin{document}

%=================================================================
% Full title of the paper (Capitalized)
\title{Hidden Pseudogap and Excitation Spectra in a Strongly Coupled Two-Band Superfluid/Superconductor}

% Author Orchid ID: enter ID or remove command
%\newcommand{\orcidauthorA}{0000-0001-5247-7116} % Add \orcidA{} behind the author's name
%\newcommand{\orcidauthorB}{0000-0001-8295-805X} % Add \orcidB{} behind the author's name
%\newcommand{\orcidauthorC}{0000-0002-4914-4975}

% Authors, for the paper (add full first names)
\author{Hiroyuki Tajima$^{1}$, Pierbiagio Pieri$^{2,3}$, and Andrea Perali$^{4}$}

% Authors, for metadata in PDF
%\AuthorNames{Firstname Lastname, Firstname Lastname and Firstname Lastname}

% Affiliations / Addresses (Add [1] after \address if there is only one affiliation.)
\affiliation{%
$^{1}$Department of Mathematics and Physics, Kochi University, Kochi 780-8520, Japan; htajima@kochi-u.ac.jp\\
$^{2}$  Dipartimento di Fisica e Astronomia, Universit\`{a} di Bologna, I-40127 Bologna, Italy; pierbiagio.pieri@unibo.it\\
$^{3}$ INFN, Sezione di Bologna, I-40127 Bologna (BO), Italy\\
$^{4}$ School of Pharmacy, Physics Unit, Universit\`{a} di Camerino, 62032 Camerino (MC), Italy; andrea.perali@unicam.it
}

% Contact information of the corresponding author
%\corres{Correspondence: }

% Current address and/or shared authorship
%\firstnote{Current address: Affiliation 3} 
%\secondnote{These authors contributed equally to this work.}
% The commands \thirdnote{} till \eighthnote{} are available for further notes

%\simplesumm{} % Simple summary

%\conference{} % An extended version of a conference paper

% Abstract (Do not insert blank lines, i.e. \\) 
\begin{abstract}
We investigate single-particle excitation properties in the normal state of a two-band superconductor or superfluid throughout the Bardeen-Cooper-Schrieffer (BCS) to Bose-Einstein-condensation (BEC) crossover, within the many-body T-matrix approximation for multi-channel pairing fluctuations. We address the single-particle density of states and the spectral functions consisting of two contributions associated with a waekly interacting deep band and a strongly interacting shallow band, relevant for iron-based multiband superconductors and multicomponent fermionic superfluids. We show how the pseudogap state in the shallow band is hidden by the deep band contribution throughout the two-band BCS-BEC crossover. 
Our results could explain  the missing pseudogap in recent scanning tunneling microscopy experiments in FeSe superconductors.
\end{abstract}
\maketitle
% Keywords
%\keyword{multi-band superconductivity; BCS-BEC crossover; ultracold Fermi gases }

% The fields PACS, MSC, and JEL may be left empty or commented out if not applicable
%\PACS{J0101}
%\MSC{}
%\JEL{}

%%%%%%%%%%%%%%%%%%%%%%%%%%%%%%%%%%%%%%%%%%
% Only for the journal Diversity
%\LSID{\url{http://}}

%%%%%%%%%%%%%%%%%%%%%%%%%%%%%%%%%%%%%%%%%%
% Only for the journal Applied Sciences:
%\featuredapplication{Authors are encouraged to provide a concise description of the specific application or a potential application of the work. This section is not mandatory.}
%%%%%%%%%%%%%%%%%%%%%%%%%%%%%%%%%%%%%%%%%%

%%%%%%%%%%%%%%%%%%%%%%%%%%%%%%%%%%%%%%%%%%
% Only for the journal Data:
%\dataset{DOI number or link to the deposited data set in cases where the data set is published or set to be published separately. If the data set is submitted and will be published as a supplement to this paper in the journal Data, this field will be filled by the editors of the journal. In this case, please make sure to submit the data set as a supplement when entering your manuscript into our manuscript editorial system.}

%\datasetlicense{license under which the data set is made available (CC0, CC-BY, CC-BY-SA, CC-BY-NC, etc.)}

%%%%%%%%%%%%%%%%%%%%%%%%%%%%%%%%%%%%%%%%%%
% Only for the journal Toxins
%\keycontribution{The breakthroughs or highlights of the manuscript. Authors can write one or two sentences to describe the most important part of the paper.}

%\setcounter{secnumdepth}{4}
%%%%%%%%%%%%%%%%%%%%%%%%%%%%%%%%%%%%%%%%%%
%%%%%%%%%%%%%%%%%%%%%%%%%%%%%%%%%%%%%%%%%%

%%%%%%%%%%%%%%%%%%%%%%%%%%%%%%%%%%%%%%%%%%
\section{Introduction}
Recently, the Bardeen-Cooper-Schrieffer to Bose-Einstein-condensation (BCS-BEC) crossover, where a weakly-interacting BCS state continuously changes to BEC of tightly bound molecules with increasing the attractive interaction~\cite{Eagles,Leggett,NSR,SadeMelo}, has gathered much attention due to its realization in ultracold Fermi gases~\cite{Regal,Zwierlein}.
Moreover, such a crossover phenomenon has been confirmed experimentally in multiband iron-based  superconductors, such as the iron-chalcogenides family FeSe~\cite{Lubashevsky,Kasahara1,Kasahara2,Rinott,Hashimoto}.
\par
Thanks to these experimental progress, the two-band BCS-BEC crossover theory has been of particular interest in condensed matter and ultracold atomic physics~\cite{Valletta,Innocenti,Guidini,He,He2,Chubukov,Mondal,Wolf,Iskin,Yerin,Aoki}.
In particular, the FeSe superconductors involve a multi-band configuration which plays a crucial role for the BCS-BEC crossover.
Moreover, the most fundamental two-band model for multichannel pairing proposed by Suhl, Matthias and Walker~\cite{SMW} has been realized in Yb ultracold Fermi gases near an orbital Feshbach resonance~\cite{Zhang,Pagano,Hofer}.
\par
One of the exciting topics in the two-band BCS-BEC crossover is the existence of pseudogaps in the single-particle excitation in the normal state~(for reviews discussing the pseudogap, see~ \cite{Mueller,Strinati,Ohashi}).
While several experiments for FeSe report signals of pseudogaps and preformed Cooper pairs~\cite{Gati,Seo2019,Kang2020,Solovjov},
a recent scanning tunneling spectroscopy (STS) measurement did not observe a pseudogap behavior even in the crossover regime of the BCS-BEC crossover~\cite{Hanaguri}. In addition, a torque magnetometry experiment in the same system indicates weak pairing fluctuations~\cite{Takahashi}.  
Theoretically, the screening of pairing fluctuations originating from the two-band configuration with different pairing strengths has been reported~\cite{Salasnich,Saraiva,Tajima1,Tajima2,Tajima3}, but the perfect screening observed in the experiment~\cite{Hanaguri} may require a further mechanism for suppressing the pseudogap.  
\par
In this article, we resolve this complicated phenomenology by calculating the single-particle density of states and spectral function throughout the two-band BCS-BEC crossover. We adopt the many-body T-matrix approximation (TMA) for multi-channel pairing fluctuations in the normal state.
We show that the pseudogap occurring in the strongly coupled shallow band is masked by the contribution from the deep band in the total (i.e., summed over the bands) density of state, which is measured in the STS experiment.
This masking effect becomes remarkable in the strong-coupling regime for the shallow band due to the overlap of spectral weights in each band.
On the other hand, we show that the total spectral function relevant for the angular-resolved-photoemission spectroscopy (ARPES) clearly reflects the pseudogap features in the strongly coupled shallow band.

\section{Formalism}
We consider a three-dimensional two-band model for attractive fermions described by
\begin{align}
H=\sum_{\bm{k},\sigma,j}\xi_{\bm{k}j}c_{\bm{k}\sigma j}^\dag c_{\bm{k}\sigma j}+\sum_{j_1,j_2}\sum_{\bm{q}}V_{j_1 j_2}b_{\bm{q}j_1}^\dag b_{\bm{q} j_2},
\end{align}
where $c_{\bm{k}\sigma j}$ is the annihilation operator of a fermion with the momentum $\bm{k}$, spin $\sigma=\up,\dwn$ and the band index $j$ (where $j=1$ and $j=2$ denote the indices of deep and shallow bands, respectively) and
$\xi_{\bm{k}j}=k^2/(2m_j)-\mu+E_0\delta_{j,2}$ is the single-particle dispersion in the $j$-band, measured from the chemical potential $\mu$ with the energy separation $E_0$ between the two bands. For simplicity, we take the same effective mass $m=m_1=m_2$ in each band.
We define a pair-annihilation operator  
\begin{align}
b_{\bm{q} j}=\sum_{\bm{k}}c_{-\bm{k}+\bm{q}/2\dwn j} c_{\bm{k}+\bm{q}/2\up j}.
\end{align}

We employ a contact-type interaction.
Specifically, the intraband couplings $V_{11}$ and $V_{22}$ are expressed in terms of 
the corresponding scattering lengths $a_{11}$ and $a_{22}$ as~\cite{Iskin}
\begin{align}
\frac{m}{4\pi a_{jj}}=\frac{1}{V_{jj}}+\sum_{\bm{k}}^{|\bm{k}|\leq k_0}\frac{m}{k^2},    
\end{align}
where the momentum cutoff $k_0$ is taken much larger
than all other momentum scales. 

\par
Superconducting pair-fluctuation effects are incorporated by the two-channel $T$-matrix~\cite{Tajima2,Tajima3}
\begin{align}
T_{jj}(\bm{q},i\nu_{\ell})=\frac{V_{jj}\left[1+V_{\bar{j}\bar{j}}\chi_{\bar{j}\bar{j}}(\bm{q},i\nu_{\ell})\right]-V_{12}V_{21}\chi_{\bar{j}\bar{j}}(\bm{q},i\nu_\ell)}{\left[1+V_{jj}\chi_{jj}(\bm{q},i\nu_{\ell})\right]\left[1+V_{\bar{j}\bar{j}}\chi_{\bar{j}\bar{j}}(\bm{q},i\nu_{\ell})\right]-V_{12}V_{21}\chi_{\bar{j}\bar{j}}(\bm{q},i\nu_\ell)\chi_{jj}(\bm{q},i\nu_\ell)},
\end{align}
where $\bar{j}$ denotes the other band with respect to band $j$ and
\begin{align}
\label{eq:chi}
\chi_{jj}(\bm{q},i\nu_\ell)=\sum_{\bm{k}}\frac{1-f\left(\xi_{\bm{k}+\bm{q}/2 \up j}\right)-f\left(\xi_{-\bm{k}+\bm{q}/2 \dwn j}\right)}{i\nu_\ell-\xi_{\bm{k}+\bm{q}/2 \up j}-\xi_{-\bm{k}+\bm{q}/2 \dwn j}}    
\end{align}
is the lowest order particle-particle correlation function.
$\nu_\ell=2\ell \pi T$ is the bosonic Matsubara frequency.
In Eq.~(\ref{eq:chi}), $f(x)=\left(e^{x/T}+1\right)^{-1}$ is the Fermi-Dirac distribution function.
In the two-channel $T$-matrix approach, 
the fermionic self-energy is of the form 
\begin{align}
\Sigma_{j}(\bm{k},i\omega_s)=T\sum_{\bm{q}}\sum_{\ell}T_{jj}(\bm{q},i\nu_\ell)G_{j}^{(0)}(\bm{q}-\bm{k},i\nu_\ell-i\omega_n),
\end{align}
where $G_{j}^{(0)}(\bm{k},i\omega_s)=\left(i\omega_s-\xi_{\bm{k} j}\right)^{-1}$ is the bare electron propagator with the fermionic Matsubara frequency $\omega_s=(2s+1)\pi T$.
The dressed propagator $G_{j}(\bm{k},i\omega_s)$ obeys the Dyson's equation
\begin{align}
G_{j}(\bm{k},i\omega_s)=G_{j}^{(0)}(\bm{k},i\omega_s)+G_{j}^{(0)}(\bm{k},i\omega_s)\Sigma_{j}(\bm{k},i\omega_s)G_{j}(\bm{k},i\omega_s).
\end{align}
\par
In the STS experiment, one observes the tunneling current $I$ occuring via the tunneling Hamiltonian~\cite{Fischer2007}
\begin{align}
H_{\rm T}=\sum_{j}\sum_{\bm{k},\bm{k}'}\sum_{\sigma,\sigma'}\left[t_j c_{\bm{k} \sigma j}^\dag c_{\bm{k}' \sigma' 0}+{\rm h.c.}\right],
\end{align}
where $c_{\bm{k}'\sigma' 0}$ denotes the annihilation operator of an electron in the weakly-coupled normal metal connected to the sample.
For simplicity, we consider the momentum-, spin-, and band-independent tunneling amplitudes $t=t_1=t_2$.
The tunneling current is obtained as
\begin{align}
\label{eq:I}
I=2\pi e\int d\omega \left[f(\omega-eV)-f(\omega)\right]
\sum_{j}\sum_{\bm{k},\bm{k}'}%\sum_{\sigma,\sigma'}
|t|^2 A_j(\bm{k},\omega-eV)A_{\rm  r}(\bm{k}',\omega),
\end{align}
where $V$ is the bias voltage and $e$ is an electron charge.
In Eq.~(\ref{eq:I}), $A_j(\bm{k},\omega)$ is the spectral function given by
\begin{align}
A_j(\bm{k},\omega)=-\frac{1}{\pi}{\rm Im}G_{j}(\bm{k},i\omega_s\rightarrow \omega+i\delta),
\end{align}
where $\delta$ is an infinitesimal small positive number to generate the retarded Green's function for real frequencies.
$A_{\rm r}(\bm{k}',\omega)$ is the spectral function in the reference normal metal.
At sufficiently low temperature, we obtain the differential conductance
\begin{align}
    \frac{dI}{dV}=2\pi e|t|^2N_{\rm r}(0)N(eV),
\end{align}
where
\begin{align}
    N(\omega)=\sum_{j}\sum_{\bm{k}} A_j(\bm{k},\omega)
\end{align}
is the total density of states. $N_{\rm r}(0)$ is the density of states at the Fermi level in the reference metal in the normal state.
In this way, one can observe if a pseudogap opens in the total density of states.

In this article, we examine $N(\omega)$ at the superconducting (or superfluid) critical temperature $T_{\rm c}$ identified by the Thouless criterion given by
\begin{align}
    \left[1+V_{11}\chi_{11}(\bm{0},0)\right]\left[1+V_{22}\chi_{22}(\bm{0},0)\right]-V_{12}V_{21}\chi_{11}(\bm{0},0)\chi_{22}(\bm{0},0)=0.
\end{align}
The chemical potential $\mu$ is determined by the density equation
\begin{align}
    n&=2T\sum_{j}\sum_{\bm{k}}\sum_{s}G_j(\bm{k},i\omega_s).
\end{align}
Note that one obtains $n=n_1+n_2$ with $n_j=\frac{k_{{\rm F},j}^3}{3\pi^2}$ in the absence of interactions at $T=0$, where $k_{{\rm F},1}=\sqrt{2mE_{\rm F,1}}$ and $k_{{\rm F},2}=\sqrt{2mE_{\rm F,2}}\equiv\sqrt{2m(E_{\rm F,1}-E_0)}$ are the band Fermi momenta ($E_{{\rm F},j}$ is the band Fermi energy). We take $E_0=\frac{3}{5}E_{\rm F,1}$ such that the two deep and shallow (occupied) bands are overlapped.
We choose the dimensionless coupling parameter in the deep band in the weak-coupling regime as $(k_{\rm F,1}a_{11})^{-1}=-2$, while the coupling parameter in the shallow band $(k_{\rm F,2}a_{22})^{-1}$ is tuned throughout the BCS-BEC crossover ($-1\lesssim (k_{\rm F,2}a_{22})^{-1}\lesssim 1$).   
The dimensionless interband pair-exchange coupling is given by $\tilde{V}_{12}=\tilde{V}_{21}=U_{12}(k_0/k_{\rm F})^2n/E_{\rm F}$ where $k_{\rm F}=(3\pi n)^{1/3}$ and $E_{\rm F}=k_{\rm F}^2/(2m)$ are the Fermi wave-vector and the Fermi energy for the total density~$n$.
We take $k_0=100k_{\rm F}$ which is a sufficiently large wave-vector cutoff compared to all other momentum scales.
\section{Results}
\begin{figure}[t]
\centering
\includegraphics[width=8 cm]{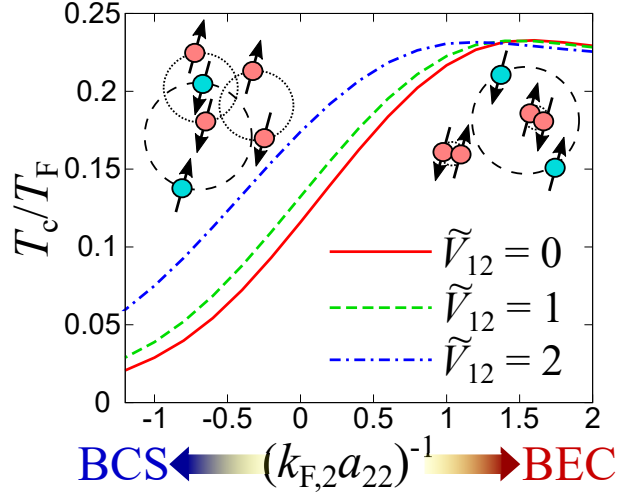}
\caption{The superconducting (superfluid) critical temperature $T_{\rm c}$ throughout the two-band BCS-BEC crossover at $(k_{\rm F,1}a_{11})^{-1}=-2$ with two cases of vanishing and finite pair-exchange couplings $\tilde{V}_{12}=0$, $\tilde{V}_{12}=1$, and $\tilde{V}_{12}=2$.
$T_{\rm F}=k_{\rm F}^2/(2m)$ is the Fermi temperature for the total number density $n$. A pictorial sketch of the two kind of Cooper pairs is given for illustration in the top part of the figure: the Cooper pairs in the shallow band (light red circles) undergo the BCS-BEC crossover, while Cooper pairs in the deep band (light blue circles) remains large, in the BCS regime.}
\label{fig1}
\end{figure}   
First, in Fig.~\ref{fig1} we show the evolution of the critical temperature $T_{\rm c}$ across the BCS-BEC crossover with increasing $(k_{\rm F,2}a_{22})^{-1}$ for three cases $\tilde{V}_{12}=0$, $\tilde{V}_{12}=1$, and $\tilde{V}_{12}=2$. 
In the weak-coupling BCS side $(k_{\rm F,2}a_{22})^{-1}\lesssim 0$, $T_{\rm c}$ exponentially increases as $\sim {\rm exp}\left(\frac{\pi}{2k_{\rm F,2}a_{22}}\right)$. 
The finite pair-exchange coupling $\tilde{V}_{12}$ gives an enhancement of $T_{\rm c}$.
In the strong-coupling BEC side $(k_{\rm F,2}a_{22})^{-1}\gesim 0$,
$T_{\rm c}$ approaches to the Bose-Einstein condensation temperature $T_{\rm BEC}$ of tightly bound molecules given by~\cite{Strinati,Ohashi}
\begin{align}
    T_{\rm BEC}&=\frac{\pi}{m}\left(\frac{n}{\zeta(3/2)}\right)^{\frac{2}{3}}\cr
    &\simeq0.218T_{\rm F},
\end{align}
where $\zeta(3/2)\simeq 2.612$ is the Riemann zeta function.
This indicates that all the particles in both bands form molecular condensates in the strong-coupling limit.
Although the nonzero pair-exchange coupling $\tilde{V}_{12}$ does not give qualitative effects on $T_{\rm c}$ in this regime [$(k_{\rm F,2}a_{22})^{-1}\gesim 1$],
the coexistence of large Cooper pairs and small molecules has been discussed within the mean-field~\cite{Yerin}, NSR~\cite{Tajima1}, and TMA~\cite{Tajima2,Tajima3} approaches.
With increasing $V_{12}$, the two-band system undergoes the BCS-BEC crossover even in the case of weak intraband couplings [$(k_{\rm F,2}a_{22})^{-1}\lesssim 1$].
In this way, one can find that the BCS-BEC crossover is realized by increasing the interaction strength in the shallow band $(k_{\rm F,2}a_{\rm 22})^{-1}$ in the present two-band model.

\begin{figure}[t]
\centering
\includegraphics[width=13 cm]{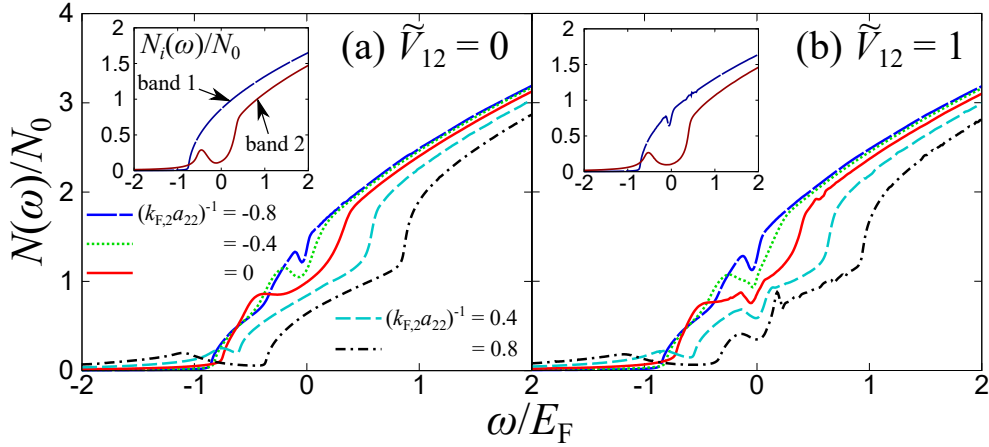}
\caption{Total density of states $N(\omega)$ relevant for the scanning tunneling spectroscopy in the two-band BCS-BEC crossover superconductor [(a) $\tilde{V}_{12}=0$, (b) $\tilde{V}_{12}=1$].The inset shows the band-selective density of states $N_i(\omega)$ at $(k_{\rm F,2}a_{22})^{-1}=0$.
$N_0=mk_{\rm F}^2/(2\pi^2)$ is the non-interacting density of states associated with the total number density $n$ at $T=0$.}
\label{fig2}
\end{figure}   
In Fig.~\ref{fig2}(a) we show the evolution of the total density of states $N(\omega)$ throughout the two-band BCS-BEC crossover with vanishing pair-exchange coupling at $T=T_{\rm c}$.
In the weak-coupling side for the shallow band [e.g. $(k_{\rm F,2}a_{22})^{-1}\leq -0.4$ in Fig.~\ref{fig2}(a)], one can see the pseudogap around the Fermi level $\omega=0$, which is small but with a relatively sharp dip structure.
On the other hand, at unitarity [crossover regime, $(k_{\rm F,2}a_{22})^{-1}=0$] in the shallow band, the pseudogap is somehow hidden.
Moreover, in the strong-coupling side of the BCS-BEC crossover [e.g. $(k_{\rm F,2}a_{22})^{-1}\geq 0.4$ in Fig.~\ref{fig2}(a)], $N(\omega)$ shows a non-monotonic structure, but not the fully gapped density of states which can be found in the single-band counterpart. 
To understand these behaviors, we examine the band-selective density of states given by
\begin{align}
 N_j(\omega)=\sum_{\bm{k}}A_j(\bm{k},\omega).   
\end{align}
In the inset of Fig.~\ref{fig2}, $N_j(\omega)$
for each band is plotted at $(k_{\rm F,2}a_{22})^{-1}=0$, corresponding to the crossover regime in the shallow band.
$N_{2}(\omega)$ clearly exhibits the pseudogap behavior (dip structure around $\omega=0$) in the shallow band ($j=2$) due to the strong pairing fluctuations associated with $V_{22}$. 
However, the deep band ($j=1$) shows the square-root behavior $N_{1}(\omega)\propto \sqrt{(\omega+\mu)}$ without the pseudogap signature in the case of $\tilde{V}_{12}=0$ because the intraband coupling is kept weak. 
In this regard, the pseudogap structure in the total $N(\omega)$ originating from $N_{2}(\omega)$ is hidden by the square-root contribution of $N_1(\omega)$.
Such a situation occurs for larger intraband coupling in the shallow band [e.g. $(k_{\rm F,2}a_{22})^{-1}=0.4$ and $0.8$ in Fig.~\ref{fig2}(a)].
On the other hand, in the case of a finite interband pair-exchange coupling $\tilde{V}_{12}=1$, shown in Fig.~\ref{fig2}(b), 
one can find a small pseudogap around $\omega=0$ in $N(\omega)$ even in the strong-coupling regime.
Furthermore, $N(\omega)$ exhibits a large flattened region around the Fermi level $(\omega=0)$.
These features can also be understood from the partial density of states $N_i(\omega)$ as shown in the inset for $(k_{\rm F,2}a_{22})^{-1}=0$.
The pair-exchange process associated with finite $V_{12}$ induces the pseudogap even in the weakly interacting deep band ($j=1$).
Hence, one can find two pseudogaps with different sizes in the two bands.
The resulting total density of states $N(\omega)$ exhibits the small pseudogap originating from $N_1(\omega)$ throughout the BCS-BEC crossover.
On the other hand, the large pseudogap in $N_2(\omega)$ is hidden by the contribution of the sizable spectral weight of $N_1(\omega)$.
\par
\begin{figure}[t]
\centering
\includegraphics[width=14.5cm]{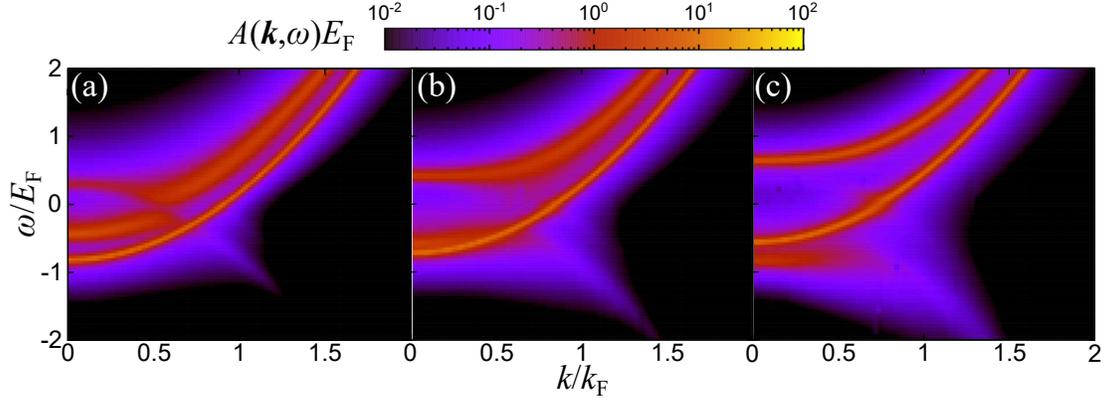}
\caption{The total spectral function $A(\bm{k},\omega)$, which is relevant for spectroscopic measurements, at (a)$(k_{\rm F,2}a_{22})=-0.4$, (b) $(k_{\rm F,2}a_{22})^{-1}=0$ and (c) $(k_{\rm F,2}a_{22})^{-1}=0.4$.
In these panels, we use $\tilde{V}_{12}=1$.
}
\label{fig3}
\end{figure}

Finally, we discuss our results for the total spectral function $A(\bm{k},\omega)$ defined as the sum of the two single-band contributions
\begin{align}
    A(\bm{k},\omega)=\sum_{j}A_{j}(\bm{k},\omega).
\end{align}
This is the quantity measured by ARPES experiments.
Figure~\ref{fig3} shows $A(\bm{k},\omega)$ with $\tilde{V}_{12}=1$, where (a) $(k_{\rm F,2}a_{22})^{-1}=-0.4$, (b) $(k_{\rm F,2}a_{22})^{-1}=0$, and (c) $(k_{\rm F,2}a_{22})^{-1}=0.4$.
Since the deep band is in the weak-coupling regime,
the dispersion originating from the deep band is close to the non-interacting counterpart given by $\omega=\xi_{\bm{k},1}$.
The pseudogap feature associated to the particle-hole mixing around $\omega=0$ is found to be weak in the deep band.
On the other hand, the shallow band exhibits the so-called Bogoliubov-like dispersion, showing the characteristic back-bending of the dispersion for large wave-vectors, given by 
\begin{align}
\omega&=\pm\sqrt{\xi_{\bm{k}_2}^2+\Delta_{\rm pg,2}^2}
\end{align}
where $\Delta_{\rm pg,2}$ is the pseudogap energy scale induced by strong pairing fluctuations.
The back-bending curve in the large wave-vector region ($p\gesim k_{\rm F}$) is one of the characteristic features for the pseudogap in the angular resolved photoemission spectroscopy of ultracold Fermi gases and strongly coupled superconductors \cite{Perali2002,Gaebler2010,Perali2011,Palestini2012}.
Since this curve is not hidden by the contribution from the deep band, it can be regarded as the signature of the pseudogap
even in the present two-band system.
For the case of strong intraband coupling $(k_{\rm F,2}a_{22})^{-1}=0.4$, the pseudogap size becomes large and the lower branch of the Bogoliubov dispersion overlaps with the deep band dispersion.
This result indicates that the tightly bound molecules in the shallow band starts dominating the system even in the presence of the cold deep band due to the very strong intraband coupling in the shallow band.

%%%%%%%%%%%%%%%%%%%%%%%%%%%%%%%%%%%%%%%%%%
\section{Conclusions}
We have investigated single-particle excitation spectra in a two-band superfluid/superconductor throughout the BCS-BEC crossover and made connections with recent experiments reporting unexpected behavior of the pseudogap in multiband iron-based superconductors. Within a two-channel T-matix approach for pairing fluctuations, we have evaluated the spectral functions and the density of states for different coupling parameters, corresponding to strong pairing in the shallow band and weak paring in the deep band, and different pair-exchange amplitudes.
We have obtained that the pseudogap in the strongly interacting regime for the shallow band is hidden by the contribution of the weakly interacting deep band in the total density of states, which is the quantity measured by the recent STS experiments in FeSe superconductors.
On the other hand, the single-particle spectral function consisting of contributions from the two bands, which is relevant to the ARPES measurement, clearly exhibits the signature of the pseudogaps, that is, the Bogoliubov back-bending dispersions.
We emphasize that these non-trivial features for the pseudogaps are unique of a two-band fermionic system in which pair-fluctuations interfere in a complex manner, originating screening or amplification phenomena of superfluid/superconductor fluctuations which are absent in the single-band counterpart.

%=====================================
% References, variant A: internal bibliography
%=====================================
\acknowledgments
Interesting discussions with Yuriy Yerin are acknowledged. H. T. is grateful for the hospitality of the Physics Division at the University of Camerino.
H. T. was supported by Grant-in-Aid for JSPS fellows
(No.17J03975) and for Scientific Research from JSPS
(No.18H05406).

%%%%%%%%%%%%%%%%%%%%%%%%%%%%%%%%%%%%%%%%%%

%\reftitle{References}

% The following MDPI journals use author-date citation: Arts, Econometrics, Economies, Genealogy, Humanities, IJFS, JRFM, Laws, Religions, Risks, Social Sciences. For those journals, please follow the formatting guidelines on http://www.mdpi.com/authors/references
% To cite two works by the same author: \citeauthor{ref-journal-1a} (\citeyear{ref-journal-1a}, \citeyear{ref-journal-1b}). This produces: Whittaker (1967, 1975)
% To cite two works by the same author with specific pages: \citeauthor{ref-journal-3a} (\citeyear{ref-journal-3a}, p. 328; \citeyear{ref-journal-3b}, p.475). This produces: Wong (1999, p. 328; 2000, p. 475)

%=====================================
% References, variant B: external bibliography
%=====================================
%\externalbibliography{yes}
%\bibliography{your_external_BibTeX_file}

\begin{thebibliography}{999}
\bibitem{Eagles} Eagles, D.M.
Possible pairing without superconductivity at low carrier concentrations in bulk and thin-film superconducting semiconductor.
{\em Phys. Rev.} {\bf 1969}, {\em 186}, 456.

\bibitem{Leggett} Leggett, A.J.
Diatomic molecules and Cooper pairs.
In {\em Modern Trends in the Theory of Condensed Matter};
\mbox{Peralski, A.,} Przystawa, Eds.; Springer: Berlin, Germany, 1980.

\bibitem{NSR}
Nozi\`eres, P.; Schmitt-Rink, S.
Bose condensation in an attractive fermion gas: From weak to strong coupling superconductivity. 
{\em J. Low Temp. Phys.} {\bf 1985}, {\em 59}, 195--211.

\bibitem{SadeMelo}
S\'{a} de Melo, C.A.R.; Randeria, M.; Engelbrecht, J.R.
Crossover from BCS to Bose Superconductivity: Transition Temperature and Time-Dependent Ginzburg-Landau Theory.
{\em Phys. Rev. Lett.} {\bf 1993}, {\em 71}, 3202.


\bibitem{Regal}
Regal, C.A.; Greiner, M.; Jin, D.S.
Observation of Resonance Condensation of Fermionic Atom Pairs.
{\em Phys.~Rev.~Lett.} {\bf 2004}, {\em 92}, 040403.

\bibitem{Zwierlein}
Zwierlein, M.W.; Stan, C.A.; Schunck, C.H.; Raupach, S.M.F.; Kerman, A.J.; Ketterle, W.
Condensation of Pairs of Fermionic Atoms near a Feshbach Resonance.
{\em Phys. Rev. Lett.} {\bf 2004}, {\em 92}, 120403.

\bibitem{Mueller} Mueller, E. J.
Review of pseudogaps in strongly interacting Fermi gases.
{\em Rep. Prog, Phys.} {\bf 2017}, {\em 80}, 104401.

\bibitem{Strinati}
Calvanese Strinati, G.; Pieri, P.; R\"{o}pke, G.; Schuck, P.; Urban, M.
The BCS-BEC crossover: From ultra-cold Fermi gases to nuclear systems.
{\em Phys. Rep.} {\bf 2018}, {\em 738}, 1. 

\bibitem{Ohashi}
Ohashi, Y.; Tajima, H.; van Wyk, P.
BCS-BEC crossover in cold atomic and in nuclear systems.
{\em Prog. Part. Nucl.~Phys.} {\bf 2020}, {\em 111}, 103739. 

\bibitem{Perali2002} 
Perali, A.; Pieri, P.; Strinati, G.C.; Castellani, C. Pseudogap and spectral function from superconducting
fluctuations to the bosonic limit. {\em Phys. Rev. B}  {\bf 2002}, {\em 66}, 024510. 

\bibitem{Gaebler2010}
Gaebler, J.P.; Stewart, J.T.; Drake, T.E.; Jin, D.S.; Perali, A.; Pieri, P.; Strinati, G.C. 
Observation of pseudogap behaviour in a strongly interacting Fermi gas. 
{\em Nat. Phys. }  {\bf 2010}, {\em 6}, 569. 

\bibitem{Perali2011}
Perali, A.; Palestini, F.; Pieri, P.; Strinati, G.C.; Stewart, J.T.; Gaebler, J.P.; Drake, T.E.; Jin D.S.
Evolution of the Normal State of a Strongly Interacting Fermi Gas from a Pseudogap Phase to a Molecular Bose Gas.
{\em Phys. Rev. Lett.}  {\bf 2011}, {\em 106}, 060402. 

\bibitem{Palestini2012} 
Palestini, F.; Perali, A.; Pieri P.; Strinati, G.C. 
Dispersions, weights, and widths of the single-particle spectral
function in the normal phase of a Fermi gas. 
{\em Phys. Rev. B}  {\bf 2012}, {\em 85}, 024517. 

\bibitem{Lubashevsky}
Lubashevsky, Y.; Lahoud, E.; Chashka, K.; Podolsky, D.; Kanigel, A. 
Shallow pockets and very strong coupling superconductivity in FeSe$_x$Te$_{1-x}$.
{\em Nat. Phys.} {\bf 2012}, {\em 8}, 309--312. 

\bibitem{Kasahara1} 
Kasahara, S.; Watashige, T.; Hanaguri, T.; Kohsaka,Y.; Yamashita, T.; Shimoyama, Y.; Mizukami, Y.; Endo, R.; Ikeda, H.; Aoyama, K.; et al.
Field-induced superconducting phase of FeSe in the BCS-BEC cross-over.
{\em Proc. Natl. Acad. Sci. USA} {\bf 2014}, {\em 111}, 16309.

\bibitem{Kasahara2} 
Kasahara, S.; Yamashita, T.; Shi, A.; Kobayashi, R.; Shimoyama, Y.; Watashige, T.; Ishida, K.; Terashima, T.; Wolf,~T.; Hardy, F.; et al.
Giant superconducting fluctuations in the compensated semimetal FeSe at the BCS-BEC crossover.
{\em Nat. Commun.} {\bf 2016}, {\em 7}, 12843.


\bibitem{Rinott}
 Rinott, S.; Chashka, K.B.; Ribak, A.; Rienks Emile, D.L.; Taleb-Ibrahimi, A.; Le Fevre, P.; Bertran, F.; Randeria,~M.; Kanigel, A.
 Tuning across the BCS-BEC crossover in the multiband superconductor Fe$_{1+y}$Se$_{x}$Te$_{1-x}$: An angle-resolved photoemission study.
{\em Sci. Adv.} {\bf 2017}, {\em 3}, e1602372.

\bibitem{Hashimoto}
Hashimoto, T.; Ota, Y.; Tsuzuki, A.; Nagashima, T.;
Fukushima, A.; Kasahara, S.; Matsuda, Y.; Matsuura, K.
Mizukami, Y.; Shibauchi, T.; Shin, S.; Okazaki, K.
Bose-Einstein condensation superconductivity induced by disappearance of the nematic state.
{\em Sci. Adv.} {\bf 2020}, {\em 6}, eabb9052.

\bibitem{Valletta}
Valletta, A.; Bianconi, A.; Perali, A.; Saini, N.L.
Electronic and superconducting properties
of a superlattice of quantum stripes at the atomic limit.
{\em Z. Phys. B: Condens. Matter} {\bf 1997}, {\em 104}, 707--713.

\bibitem{Iskin}
Iskin, M.; S\'{a} de Melo, C.A.R.
Two-band superfluidity from the BCS to the BEC limit.
{\em Phys. Rev. B} {\bf 2006},~{\em 74},~144517.

\bibitem{Innocenti}
Innocenti, D.; Poccia, N.; Ricci, A.; Valletta, A.; Caprara, S.; Perali, A.; Bianconi, A.
Resonant and crossover phenomena in a multiband superconductor: Tuning the chemical potential near a band edge.
{\em Phys. Rev. B} {\bf 2010},~{\em 82},~184528.


\bibitem{Guidini}
Guidini, A.; Perali, A.
Band-edge BCS-BEC crossover in a two-band superconductor:
physical properties and detection parameters.
{\em Supercond. Sci. Technol.} {\bf 2014}, {\em 27}, 124002.

\bibitem{He2}
He, L.; Hu, H.; Liu, X.-J.
Two-band description of resonant supefluidity in atomic Fermi gases.
{\em Phys. Rev. A} {\bf 2015},~{\em 91},~023622.


\bibitem{He}
He, L.; Wang, J.; Peng, S.-G.; Liu, X.-J.; Hu, H. 
Strongly correlated Fermi superfluid near an orbital Feshbach resonance: Stability, equation of state, and Leggett mode.
{\em Phys. Rev. A} {\bf 2016}, {\em 94}, 043624.

\bibitem{Chubukov}
Chubukov, A. V.; Eremin, I.; Efremov, D. V.
Superconductivity versus bound-state formation in a two-band superconductor with small Fermi energy:
Applications to Fe pnictides/chalcogenides and doped SrTiO$_3$.
{\em Phys. Rev. B} {\bf 2016}, {\bf 93}, 174516.

\bibitem{Wolf}
Wolf, S.; Vagov, A.; Shanenko, A.A.; Axt, V.M.; Perali, A.; Albino Aguiar, J.
BCS-BEC crossover induced by a shallow band: Pushing standard superconductivity types apart.
{\em Phys. Rev. B} {\bf 2017}, {\em 95}, 094521.

\bibitem{Mondal}
Mondal, S.; Inotani, D.; Ohashi, Y.
Single-particle Excitations and Strong-Coupling Effects in the BCS-BEC Crossover Regime of a Rare-Earth Fermi Gas with an Orbital Feshbach Resonance.
{\em J. Phys. Soc. Jpn.} {\bf 2018},~{\em 87},~084302.


\bibitem{Yerin}
Yerin, Y.; Tajima, H.; Pieri, P.; Perali, A.
Coexistence of giant Cooper pairs with a bosonic condensate and anomalous behavior of energy gaps in the BCS-BEC crossover of a two band superfluid Fermi gas.
{\em Phys. Rev. B} {\bf 2019}, {\em 100}, 104528.

\bibitem{Aoki}
Aoki, H.
Theoretical Possibilities for Flat Band Superconductivity.
{\em J. Supercond. Nov. Magn.} {\bf 2020}, {\em 33}, 2341-2346.

\bibitem{SMW}
Suhl, H.; Matthias, B.T.; Walker, L.R.
Bardeen--Cooper--Schrieffer Theory of Superconductivity in the Case of Overlapping Bands.
{\em Phys. Rev. Lett.} {\bf 1959}, {\em 3}, 552.

\bibitem{Zhang}
Zhang, R.; Cheng, Y.; Zhai, H.; Zhang, P.
Orbital Feshbach Resonance in Alkali-Earth Atoms.
{\em Phys. Rev. Lett.} {\bf 2015}, {\em 115}, 135301.


\bibitem{Pagano}
Pagano, G.; Mancini, M.; Cappellini, G.; Livi, L.; Sias, C.; Catani, J.; Inguscio, M.; Fallani, L.
Strongly Interacting Gas of Two-Electron Fermions at an Orbital Feshbach Reonance.
{\em Phys. Rev. Lett.} {\bf 2015}, {\em 115}, 265301.

\bibitem{Hofer}
H\"{o}fer, M.; Riegger, L.; Scazza, F.; Hofrichter, C.; Fernandes, D.R.; Parish, M.M.; Levinsen, J.; Bloch, I.; Folling, S.
Observation of an Orbital Interaction-Induced Feshbach Resonance in $^{173}$Yb.
{\em Phys. Rev. Lett.} {\bf 2015}, {\em 115}, 265302.
 
\bibitem{Salasnich}
Salasnich, L.; Shanenko, A.A.; Vagov, A.; Albino Aguiar, J.; Perali, A.
Screening of pair fluctuations in superconductors with coupled shallow and deep bands: A route to higher-temperature superconductivity.
{\em Phys. Rev. B} {\bf 2019}, {\em 100}, 064510.

\bibitem{Saraiva}
Saraiva, T.T.; Cavalcanti, P.J.F. ; Vagov, A.; Vasenko A.S.; Perali, A.; Dell’Anna, L.; Shanenko, A.A.
Multiband Material with a Quasi-1D Band as a Robust High-Temperature Superconductor.
{\em Phys. Rev. Lett.} {\bf 2020},  {\em 125}, 217003.

\bibitem{Tajima1} Tajima, H.; Yerin, Y.; Perali, A.; Pieri, P.
Enhanced critical temperature, pairing fluctuation effects, and BCS-BEC crossover in a two band Fermi gas.
{\em Phys. Rev. B} {\bf 2019}, {\em 99}, 180503(R).

\bibitem{Tajima2} Tajima, H.; Pieri, P.; Perali, A.
BCS-BEC Crossover and Pairing Fluctuations in a Two Band Superfluid/Superconductors: A $T$ Matrix Approach.
{\em Condens. Matter}, {\bf 2020}, {\em 5}, 10.

\bibitem{Tajima3} Tajima, H.; Yerin, Y.; Pieri, P.; Perali, A.
Mechanism of screening or enhancing the pseudogapp throughout the two-band Bardeen-Cooper-Schrieffer to Bose-Einstein condensate crossover.
{\em Phys. Rev. B} {\bf 2020}, {\em 102}, 220504(R).


\bibitem{Gati} 
Gati, E.; B\"{o}hmer, A.E.; Bud'ko, S.L.; Canfield, P.C.
Bulk superconductivity and role of fluctuations in the iron based superconductor FeSe at high pressures.
{\em Phys. Rev. Lett.} {\bf 2019}, {\em 123}, 167002.

\bibitem{Seo2019}
Seo, Y. I.; Choi, W. J.; Kimura S.; Kwon, Y. S. 
Evidence for a preformed Cooper pair model in the pseudogap spectra of a Ca$_{10}$(Pt$_4$As$_8$)(Fe$_2$As$_2$)$_5$ single crystal with a nodal superconducting gap.
{\em Sci. Rep.} {\bf 2019}, {\em 9}, 3987.

\bibitem{Kang2020} Kang, B. L.; Shi, M. Z.; Li, S. J.; Wang, H. H.; Zhang, Q.; Zhao, D.; Li, J.; Song, D. W. ;Zheng, L. X.; Nie, L. P.; Wu, T.; Chen, X. H.
Preformed Cooper Pairs in Layered FeSe-Based Superconductors.
{\em Phys. Rev. Lett.} {\bf 2020}, {\em 125}, 097003.

\bibitem{Solovjov} Solovjov, A. L.; Petrenko, E. V.; Omelchenko, L.V.; Nazarova, E.; Buchkov, K.; Rogacki, K.
Fluctuating Cooper pairs in FeSe at temperatures exceeding double $T_{\rm c}$.
{\em arXiv:2010.12319 [cond-mat.supr-con]} {\bf 2020}.


\bibitem{Hanaguri} 
Hanaguri, T.; Kasahara, S.; B\"{o}ker, J.; Eremin, I.; Shibauchi, T.; Matsuda, Y.
Quantum Vortex Core and Missing Pseudogap in the Multiband BCS-BEC Crossover Superconductor FeSe.
{\em Phys. Rev. Lett.} {\bf 2019}, {\em 122}, 077001.

\bibitem{Takahashi} 
Takahashi, H.; Nabeshima, F.; Ogawa, R.; Ohmichi, E.; Ohta, H.; Maeda, A.
Superconducting fluctuations in FeSe investigated by precise torque magnetometry.
{\em Phys. Rev. B} {\bf 2019}, {\em 99}, 060503(R).

\bibitem{Fischer2007}
Fischer, \O.;Kugler, M.; Maggio-Aprile, I.; Berthod, C.;
Renner, C.
Scanning tunneling spectroscopy of high-temperature superconductors.
{\em Rev. Mod. Phys.} {\bf 2007}, {\em 79}, 353.

\end{thebibliography}

%%%%%%%%%%%%%%%%%%%%%%%%%%%%%%%%%%%%%%%%%%
%% optional
%\sampleavailability{Samples of the compounds ...... are available from the authors.}

%% for journal Sci
%\reviewreports{\\
%Reviewer 1 comments and authors’ response\\
%Reviewer 2 comments and authors’ response\\
%Reviewer 3 comments and authors’ response
%}

%%%%%%%%%%%%%%%%%%%%%%%%%%%%%%%%%%%%%%%%%%
\end{document}